\documentclass[conference]{IEEEtran}
\IEEEoverridecommandlockouts
\usepackage{cite}
\usepackage{amsmath,amssymb,amsfonts}
\usepackage{algorithmic}
\usepackage{graphicx}
\usepackage{textcomp}
\usepackage{xcolor}

\usepackage{amsmath}
\usepackage{amssymb}
\usepackage{booktabs}
\usepackage{placeins}
\usepackage{float}
\usepackage{graphicx}  
\usepackage{subcaption}
\usepackage{multirow} 
\usepackage{siunitx} 
\usepackage{url}

\def\BibTeX{{\rm B\kern-.05em{\sc i\kern-.025em b}\kern-.08em
    T\kern-.1667em\lower.7ex\hbox{E}\kern-.125emX}}
\begin{document}

\title{Physical-Layer Analysis of LoRa Robustness in the Presence of Narrowband Interference 
}

\author{\IEEEauthorblockN{Jingxiang Huang}
\IEEEauthorblockA{\textit{Faculty of Computer Science} \\
\textit{Dalhousie University}\\
Halifax, Canada \\
jn661503@dal.ca}
\and
\IEEEauthorblockN{Samer Lahoud}
\IEEEauthorblockA{\textit{Faculty of Computer Science} \\
\textit{Dalhousie University}\\
Halifax, Canada \\
sml@dal.ca}
}

\maketitle

\begin{abstract}
With the rapid development of Internet of Things (IoT) technologies, the sub-GHz unlicensed spectrum is increasingly being shared by protocols such as Long Range (LoRa), Sigfox, and Long-Range Frequency-Hopping Spread Spectrum (LR-FHSS).  These protocols must coexist within the same frequency bands, leading to mutual interference. This paper investigates the physical-layer impact of two types of narrowband signals (BPSK and GMSK) on LoRa demodulation.  We employ symbol-level Monte Carlo simulations to analyse how the interference-to-noise ratio (INR) affects the symbol error rate (SER) at a given signal-to-noise ratio (SNR) and noise floor, and then compare the results with those for additive white Gaussian noise (AWGN) of equal power.  We demonstrate that modelling narrowband interference as additive white Gaussian noise (AWGN) systematically overestimates the SER of Chirp Spread Spectrum (CSS) demodulation. We also clarify the distinct impairment levels induced by AWGN and two types of narrowband interferers, and provide physical insight into the underlying mechanisms.  Finally, we fit a two-segment function for the maximum INR that ensures correct demodulation across SNRs, with one segment for low SNR and the other for high SNR.

\end{abstract}


\section{Introduction}

\subsection{Background}
Coexistence among Internet of Things (IoT) devices in unlicensed spectrum has become an important challenge, with more systems sharing the same frequency bands.  In contrast to licensed spectrum, where regulators allocate dedicated channels, unlicensed bands rely on simple rules such as duty-cycle limits and carrier sensing mechanisms (also known as Listen Before Talk, LBT). As device density grows in urban, industrial, and agricultural settings, transmissions from different technologies can overlap in time or frequency, causing interference that degrades performance for all users.

Low-Power Wide-Area Networks (LPWANs) are a family of radio technologies for battery-powered devices that need to send small amounts of data over kilometers of range. LPWANs operate in the sub-GHz unlicensed bands and employ frequency hopping or spread-spectrum modulation, extremely low duty cycles, and simple MAC protocols to minimize power consumption and cost.  

LoRa (Long Range) is a widely adopted LPWAN protocol that achieves long-range coverage by using chirp spread spectrum (CSS) at the physical layer \cite{Semtech,lora}.  Other LPWAN protocols, such as Sigfox and Long-Range Frequency-Hopping Spread Spectrum (LR-FHSS), adopt narrowband approaches in the sub-GHz unlicensed bands \cite{Sigfox,fhss}. When their transmissions overlap a LoRa channel, the interference can corrupt the demodulation process and increase the symbol error rate (SER). Understanding how different types of  narrowband interference affect LoRa demodulation is helpful for developing targeting LBT strategies and improve coexistence among diverse IoT technologies.

\subsection{Related Works}

Intra-technology interference happens when multiple LoRa devices use the same spreading factor (SF) simultaneously. The work \cite{intra1} presents a mathematical model and simulations of the interference between two LoRa symbols with the same SF when they are chirp- or phase-aligned, re-evaluating the coverage for the LoRa network. Furthermore, the work \cite{intra2} expanded on the previous work, proving that assuming full chirp-level alignment represents a worst-case scenario. From another perspective,  another research \cite{intra3} questioned the orthogonality of different spreading factors. It quantified imperfect orthogonality through simulation, which highlights that LoRa’s physical-layer orthogonality cannot prevent collisions in dense networks. 

LoRa also suffers from inter-technology interference. The impact of inter-technology interference has been measured under different scenarios. The work \cite{inter1} investigated the impact of broadband jamming and single-tone jamming and found that LoRa’s chirp spread spectrum can mitigate the effect of tone jamming, whereas broadband jamming simply raises the noise floor.  The papers \cite{inter2,inter3} used network models to investigate and simulate Sigfox interference on LoRa, examining how transmit power, distance, and the number of nodes affect LoRa’s error rate and throughput. However, neither of them discusses how the interference signal would affect the demodulation process at the physical layer. 



LoRa also supports operation in the 2.4 GHz band, where it coexists with Wi-Fi, Bluetooth Low Energy (BLE) and Long Term Evolution (LTE), all of which act as sources of broadband interference for LoRa. The work in \cite{2.4g1,2.4g2,2.4g3} studied the interference brought by Wi-Fi, BLE, and LTE at the physical layer, respectively. The authors demonstrate that LoRa signals exhibit robustness against broadband interference and further show that this resilience depends on factors such as the signal-to-interference ratio (SIR), the modulation schemes used by the interferers, and the degree of overlap between the LoRa and interference bandwidths. 

Given the various types of interference impacting LoRa network performance, detecting and avoiding interference at the physical layer is crucial for improving coexistence. Energy-based carrier sensing is a traditional approach. The work \cite{ed} highlights that the choice of sensing threshold is critical: if the threshold is too high, devices may overlook weak but valid signals; setting the threshold too low can make the device overly sensitive to random noise. Although energy-based carrier sensing is simple to implement, it cannot distinguish between noise and different types of interference, which limits its potential for more nuanced approaches. 

Many studies have examined how inter-technology interference degrades LoRa reception through measurements and simulations. However, physical-layer analyses that elucidate the mechanisms by which different interferer types uniquely affect LoRa demodulation are still lacking. Furthermore, in the context of LoRa networks, it is still unclear if white noise and interference can be treated equivalently in network-performance assessments or when designing carrier sensing schemes.

\subsection{Contributions}

The main contributions of this paper are as follows: i) We demonstrate that modeling narrowband interference as AWGN results in excessively conservative performance predictions through Monte Carlo simulations. ii) By applying the principle of stationary phase, we derive an explicit relationship between the complex envelope of interference and its impact on the CSS decision statistic. iii) Comprehensive Monte Carlo experiments for spreading factors 7-12 are condensed into a two-segment analytical fit that yields the maximum tolerable interference-to-noise power ratio (INR) as a function of signal-to-noise ratio (SNR).

\section{LoRa Demodulation Process}
In this section, we present the mathematical analysis of the LoRa demodulation process in the presence of AWGN and interference. 

Suppose the bandwidth of LoRa is denoted as \(B\), the duration of a LoRa symbol is determined by its bandwidth and spreading factor (SF) according to the following equation 
\[
T_s = \frac{2^{\mathrm{SF}}}{B}.
\]

For a SF \(\in \{7, \dots, 12\} \), each LoRa symbol encodes SF bits, yielding \( N = 2^\text{SF} \) distinct symbols. The symbol indices range from \(0\) to \(N-1\), and the symbol at index 0 is called the \emph{upchirp}, which is defined as
\begin{equation}
\sigma_0[n] = \exp\!\Bigl(j\pi\frac{n(n-N)}{N}\Bigr).
\end{equation}

According to the datasheet, when a Semtech’s LoRa chip SX1276/7/8/9 processes an incoming signal, the signal is down-converted to baseband and goes through an ADC for analog-to-digital conversion \cite{Semtech}. We assume the sampling rate equals the bandwidth of the LoRa signal.

Denote the received signal  \(\sigma[n]\) as an addition of the LoRa signal \(\sigma_s[n]\), AWGN  \(\sigma_n[n]\) and interference  \(\sigma_i[n]\) for each sampling point \(n\)
\begin{equation}
\sigma[n] = \sigma_s[n]+\sigma_n[n]+\sigma_i[n] \quad n=0,1,...,N-1.
\end{equation}

The first step of demodulation is \textit{dechirping}, which consists of multiplying the received signal by the complex conjugate of the upchirp \(\sigma_0^*[n]\) for each sampling point \(n\)
\begin{equation}
    y[n] = \sigma_0^*[n]\, \sigma[n] =  \sigma_0^*[n]( \sigma_s[n]+ \sigma_n[n]+ \sigma_i[n]).
\end{equation}

An \(N\)-point Discrete Fourier Transform (DFT) is then applied to \(y[n]\) in order to obtain the decision statistic \(Y[k]\), which is given by
\begin{equation}
Y[k] = \sum_{n=0}^{N-1} y[n]\, e^{-j\frac{2\pi}{N}kn}\quad k=0,1,...,N-1.
\end{equation}

The transmitted symbol index \(p\in\{0,1,...,N-1\}\) can be recovered by finding the DFT bin in \(Y[k]\) with the largest magnitude
\begin{equation}
\hat{p}
\;=\;
\arg\max_{0 \,\le\, k \,<\, N}\;\bigl|Y[k]\bigr|.
\label{decision}
\end{equation}

By expanding formula (4), we obtain three terms, \(Y_s[k]\), \(Y_n[k]\), and \(Y_i[k]\), corresponding to the contribution made by the LoRa signal, AWGN, and the interference, respectively
\begin{align*}
Y[k]&= \sum_{n=0}^{N-1} \sigma_0^*[n]\, \sigma_s[n]\, e^{-j\frac{2\pi}{N}kn}+\sum_{n=0}^{N-1} \sigma_0^*[n]\, \sigma_n[n]\, e^{-j\frac{2\pi}{N}kn}\\
     &+\sum_{n=0}^{N-1} \sigma_0^*[n]\, \sigma_i[n]\, e^{-j\frac{2\pi}{N}kn}\\
     &=Y_s[k]+Y_n[k]+Y_i[k].
\end{align*}

For the first term \(Y_s[k]\), the LoRa symbol with index \(p\) produces a peak at \(k=p\)
\begin{equation}
Y_s[k] = \sum_{n=0}^{N-1} \sigma_0^*[n]\, \sigma_s[n]\;e^{-j\frac{2\pi}{N}kn}
=\begin{cases}
\sqrt{P_s}N, &  k = p,\\
0, & k \neq p,
\end{cases}
\end{equation}
where \(P_s\) represents the average power of the received LoRa signal.

The second term \(Y_n[k]\) represents the contribution of AWGN. We model the noise in the time domain as a zero‐mean complex Gaussian distribution \(\sigma_n[n]\sim\mathcal{CN}(0,\frac{P_n}{2})\), where \(\frac{P_n}{2}\) denotes the variance in each dimension, and \(P_n\) denotes the average noise power (the noise floor). After dechirping and performing the DFT, \(Y_n[k]\) also follows a complex Gaussian distribution given by \(\mathcal{CN}\bigl(0,\,N\frac{P_n}{2}\bigr)\).

The term \(Y_i[k]\) corresponds to the effect of narrowband interference. Assume a narrowband signal with complex envelope \( i(t) \), whose average power is \(P_i\). Denote the difference between its centre frequency and LoRa’s centre frequency as \( \Delta f \). After sampling and down-conversion, the resulting baseband signal sampled at rate \( f_s = B \) is given by
\begin{equation}
\sigma_i[n] = i[n]  \exp\!\Bigl(j2\pi \Delta f\frac{n}{B} \Bigr), \quad n=0,\dots,N-1,
\end{equation}
where \( i[n] = i\left(\frac{n}{B}\right) \).

Then \(Y_i[k]\) is given by
\begin{equation}
Y_i[k]
= \sum_{n=0}^{N-1} \sigma_0^*[n]\, \sigma_i[n]\, e^{-j\frac{2\pi}{N}kn}
= \sum_{n=0}^{N-1} i[n]\, e^{\,jF[n;k]},
\end{equation}
with the phase term defined as
\begin{equation}
F[n;k]
= 2\pi\Delta f \frac{n}{B} - \pi\frac{n(n-N)}{N} - \frac{2\pi k}{N}n.
\label{phaseF}
\end{equation}

\begin{figure}[htbp]
\centerline{\includegraphics[width=1\linewidth]{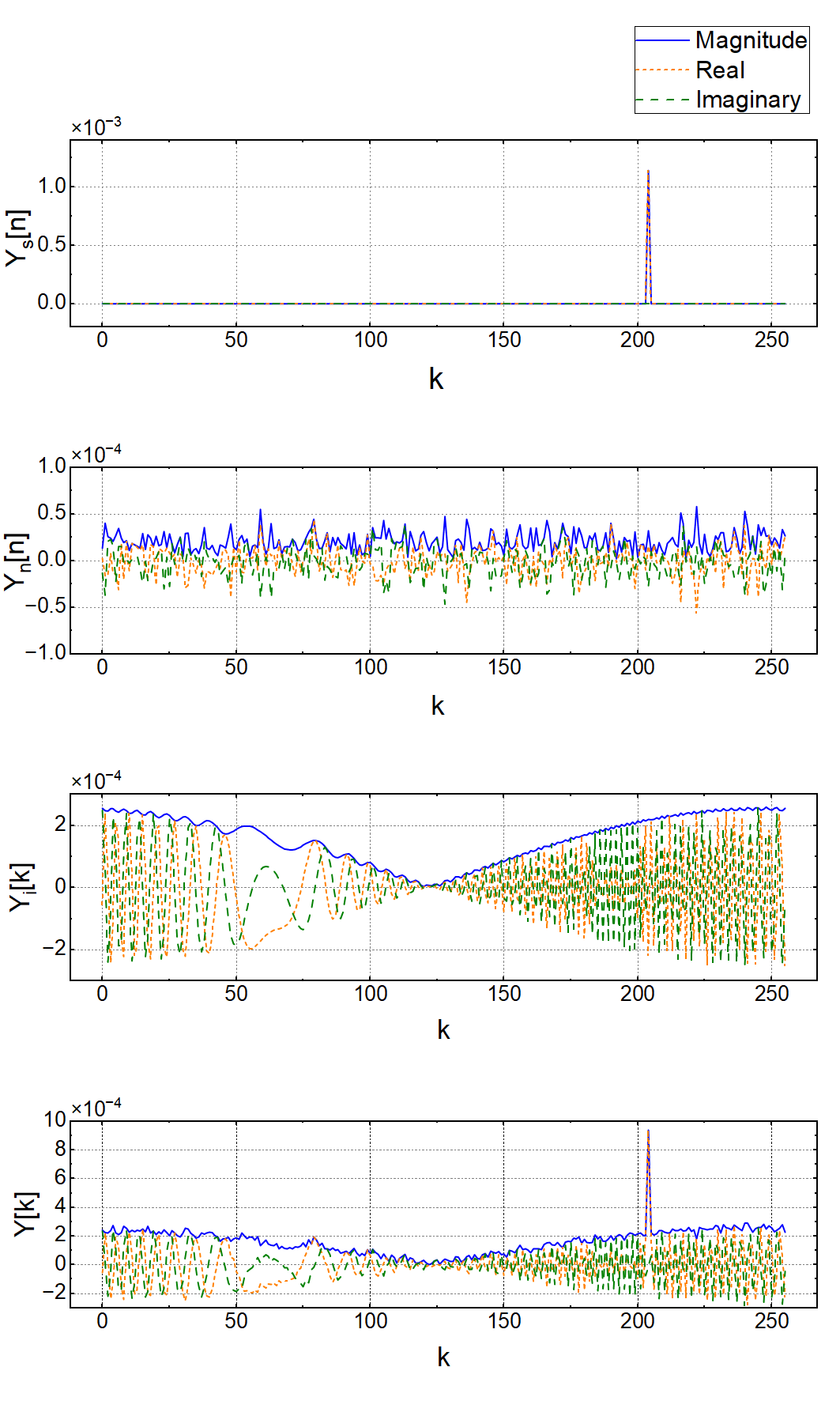}}
\caption{Frequency-domain example of \(Y_s[k]\),\(Y_n[k]\),\(Y_i[k]\) and \(Y[k]\).}
\label{example}
\end{figure}

 Fig. \ref{example} shows an example of \(Y_s[k]\),\(Y_n[k]\),\(Y_i[k]\) and \(Y[k]\), all of which are frequency-domain quantities. Based on the preceding analysis, the LoRa signal with index \( p \) produces a peak at bin \( k = p \) in the DFT output \( Y_s[k] \) while the other bins remain zero. The AWGN contributes a random complex value to every bin of  \( Y_n[k] \), and the interference also affects all bins in \( Y_i[k] \), with its impact determined by its complex envelope \(i[n]\) and frequency offset \(\Delta f\).  The sum of these three terms yields \(Y[k]\) and the symbol is recovered based on \(|Y[k]|\) according to (\ref{decision}).

\section{Simulation Setup and Result}

In this section, we evaluate the impact of narrowband interference on LoRa demodulation through Monte Carlo simulations. Based on the analysis in the previous section, we construct a symbol-level simulation framework that models the impact of the LoRa signal, AWGN, and narrowband interferers. By varying the modulation type, INR, SNR, we observe the SER under different conditions. The simulation results provide quantitative insights into the robustness of LoRa demodulation under AWGN and interference.

\begin{figure}[htbp]
\centerline{\includegraphics[width=0.8\linewidth]{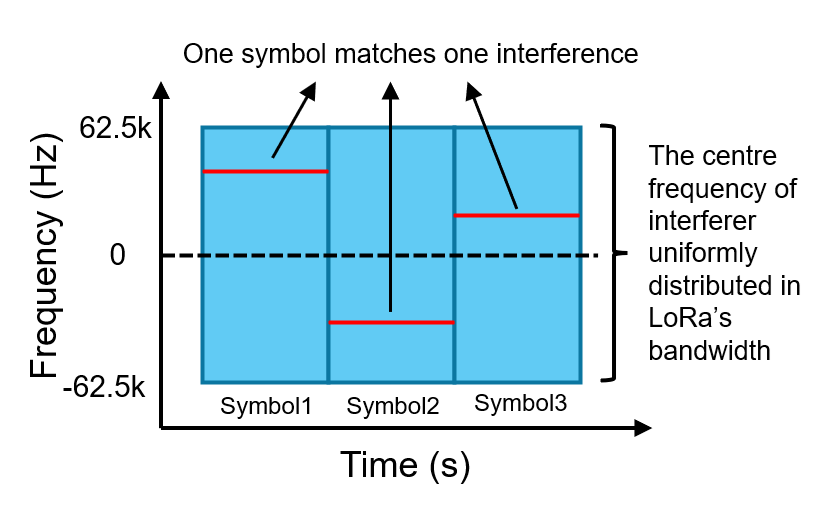}}
\caption{Illustration of Assumptions.}
\label{bandwidth}
\end{figure}

To simplify the analysis, we make several assumptions. Firstly, we assume that the bandwidth of the LoRa signal fully includes that of the narrowband interference, meaning the interference lies entirely within the LoRa bandwidth. Since interference can occur only when frequency bands overlap and narrowband Sigfox or LR-FHSS interferers occupy only a few hundred hertz, which is insignificant compared to LoRa channels spanning several hundred kilohertz, so we ignore cases of partial overlap. Secondly, the centre frequency of the interferer is uniformly distributed across the LoRa bandwidth, representing frequency-hopping applied by Sigfox and LR-FHSS. Thirdly, perfect temporal alignment is assumed, which means that each LoRa symbol is affected by exactly one narrowband interferer throughout its duration. In cases of temporal misalignment, the impact on demodulation is reduced; this will be discussed in Subsection B. Fig. \ref{bandwidth} shows the above assumptions. Also, this simulation is conducted at the symbol level; therefore, no channel coding techniques such as forward error correction, whitening, or interleaving are applied.

Two types of interference signals are considered in the simulation: BPSK and GMSK, both with a bandwidth of 600~Hz. These two modulation schemes are used in Sigfox and LR-FHSS, respectively. In practice, Sigfox uplink uses BPSK with a 600~Hz bandwidth in North America, while LR-FHSS uses GMSK with a bandwidth of 488~Hz \cite{Sigfox,fhss}. For consistency in controlled comparisons, both signals are configured with the same bandwidth (600 Hz) in the simulation. To serve as a control group, we also simulate a case where an additional AWGN with the same power as the narrowband interference is added to the original noise floor. This setup allows a direct comparison between noise and interference in terms of their impact on LoRa demodulation under equal power conditions.

In the Monte Carlo simulation, we model the three components contributing to the decision variable \(Y[k]\): the LoRa‐signal component \(Y_s[k]\), the AWGN component \(Y_n[k]\), and the narrowband‐interference component \(Y_i[k]\). For the LoRa signal, one bin in \( Y_s[k] \) is randomly selected, and its real part is set to \( \sqrt{P_s }N \), while its imaginary part and other bins remain zero, representing the correct symbol. For AWGN, we add to \(Y[k]\) a complex Gaussian noise term \(Y_n[k]\), where each bin follows \(\mathcal{CN}\!\bigl(0,\,N\frac{P_n}{2}\bigr)\). For the narrowband interference, a baseband waveform (BPSK or GMSK) is first generated and sampled at the rate of \( B \), and its average power is normalized to \( P_i \). A chunk of the interference waveform is extracted by selecting a random starting index and taking a continuous segment of samples equal in length to one LoRa symbol. A frequency offset, uniformly selected within the LoRa bandwidth, is applied to simulate a random position within the spectrum. The resulting signal is then multiplied by the conjugate of the upchirp and transformed by an \( N \)-point DFT. The resulting interference component \(Y_i[k]\) is added to \( Y[k] \) to represent the interference contribution. After the three components are combined, the bin with the maximum magnitude in \( Y[k] \) is identified. If this bin matches the index of the correct LoRa symbol, the demodulation is considered successful; otherwise, an error is occurred. 

For all simulation scenarios, the LoRa signal bandwidth is set to 125~kHz. At room temperature (\(T = 290~\text{K}\)), the corresponding thermal noise power is
\begin{equation}
N_{\text{thermal}} = k_B T B = -123~\text{dBm},
\end{equation}
where \(k_B\) is Boltzmann’s constant. Considering a typical receiver noise figure of \(NF = 6~\text{dB}\) \cite{Semtech}, the total noise floor becomes
\begin{equation}
N_0 = N_{\text{thermal}} + NF = -117~\text{dBm}.
\end{equation}

For each configuration, 10{,}000 independent symbol transmissions are simulated to ensure statistical reliability of the results.

\subsection{Symbol Error Rate of LoRa with AWGN}
\begin{figure}
\centerline{\includegraphics{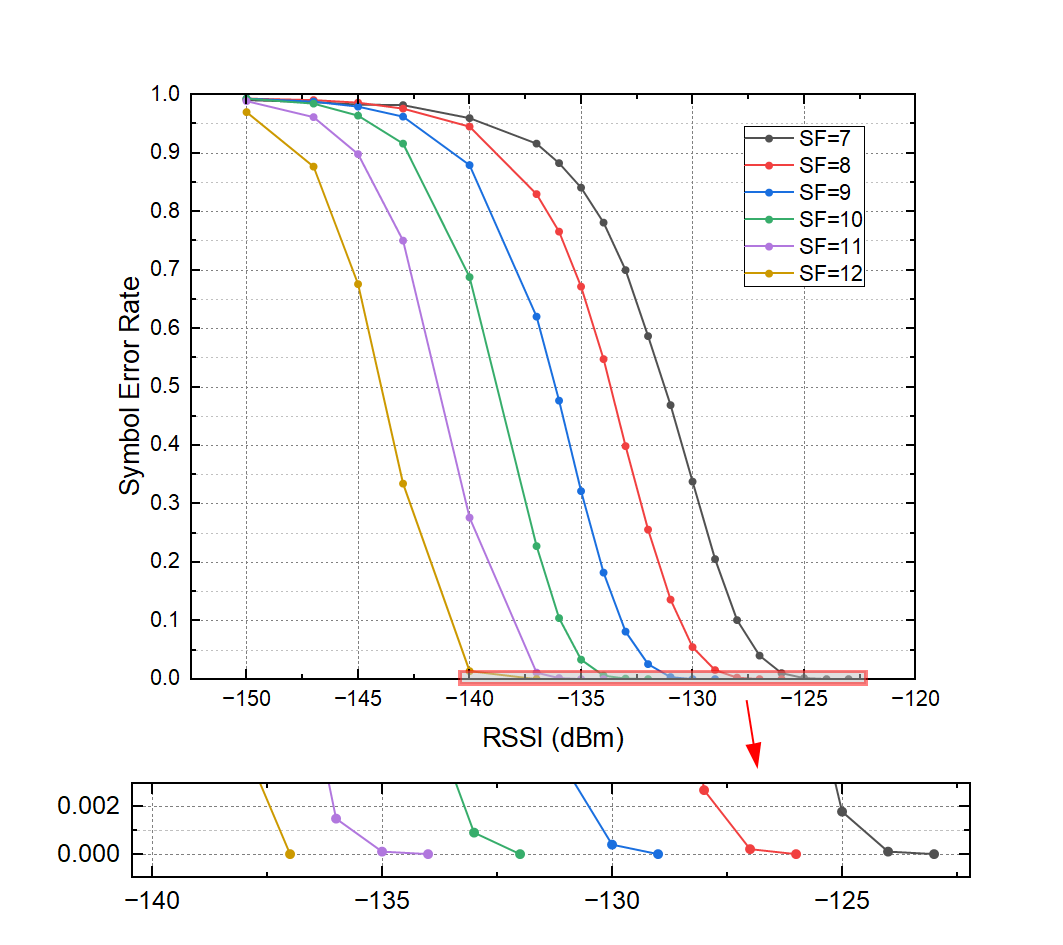}}
\caption{Symbol Error Rate versus RSSI.}
\label{fig1}
\end{figure}

We first conduct tests under noise-only conditions without interference. Fig. \ref{fig1} demonstrates the SER performance versus received signal strength (RSSI) for LoRa signals with different spreading factors. Table \ref{tab:rssi_thresholds} lists the minimum RSSI values required to achieve a zero symbol error rate, which we denote as \(R_T\). The values of \(R_T\) closely match the sensitivity figures in the LoRa datasheet \cite{Semtech}, suggesting an inherent physical-layer relationship between \(R_T\) and LoRa sensitivity.

\begin{table}[htbp]
  \centering
  \caption{Comparison of \(R_T\) values and LoRa sensitivity}
  \label{tab:rssi_thresholds}
  \begin{tabular}{c c c}
    \hline
    SF  & RSSI threshold \(R_T\) (dBm) & LoRa sensitivity (dBm) \\
    \hline
    7   & -123                         & -123                   \\
    8   & -126                         & -126                   \\
    9   & -129                         & -129                   \\
    10  & -132                         & -132                   \\
    11  & -134                         & -134.5                 \\
    12  & -137                         & -137                   \\
    \hline
  \end{tabular}
\end{table}

\subsection{Symbol Error Rate of LoRa with AWGN and Interference}
\begin{figure}[htbp]
\centerline{\includegraphics{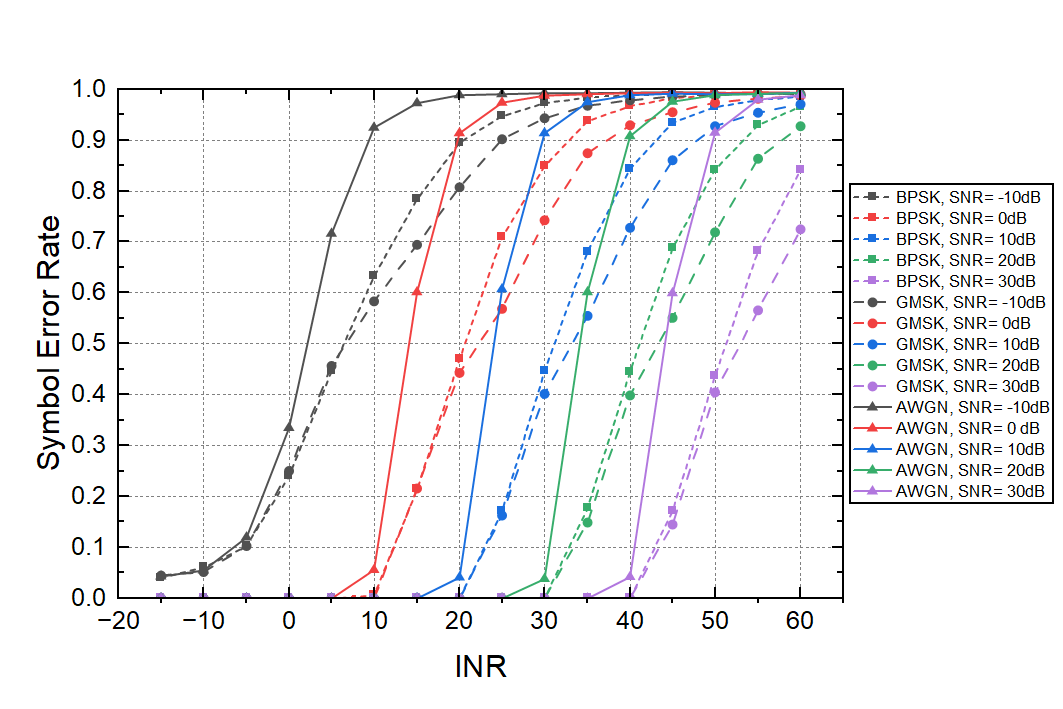}}
\caption{Curves of Symbol Error Rate versus INR under different SNR for SF7.}
\label{fig2}
\end{figure}

\begin{figure}[htbp]
\centerline{\includegraphics{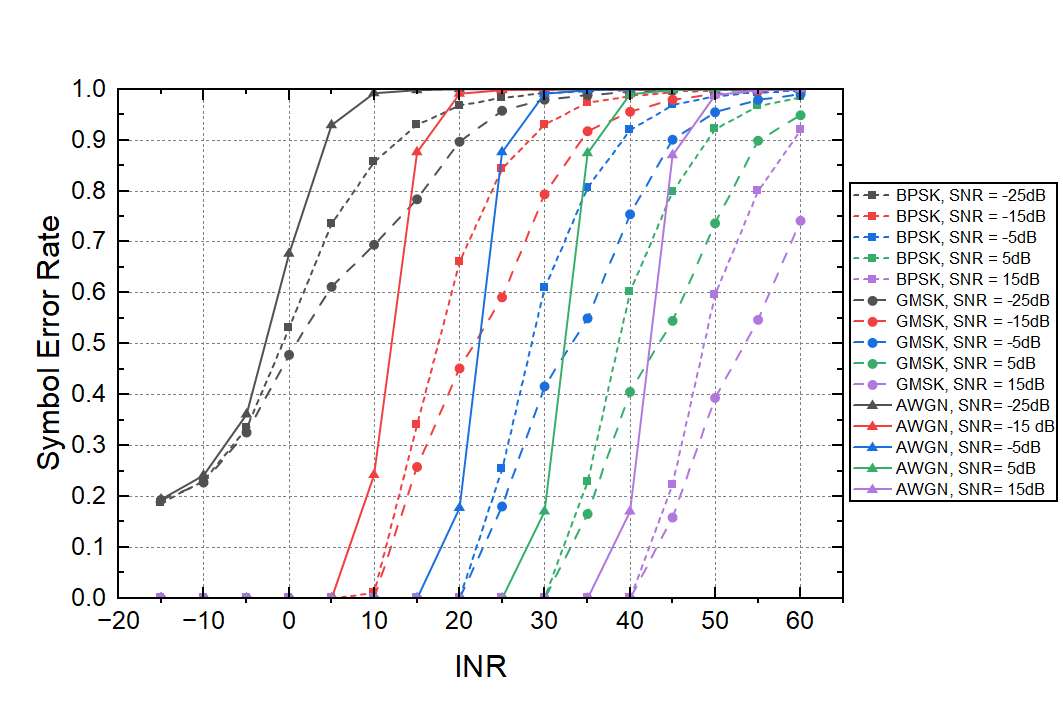}}
\caption{Curves of Symbol Error Rate versus INR under different SNR for SF12.}
\label{fig3}
\end{figure}

Fig. \ref{fig2} and \ref{fig3} show the relations between SER  and INR under different SNR for SF7 and SF12, respectively.  In the figures, AWGN refers to the control group. 

These  figures demonstrate that, given the same power level, AWGN leads to the most severe degradation in LoRa demodulation, BPSK results in moderate degradation, and GMSK causes the least impact among them. This shows that different types of interference and noise have varying impacts on LoRa demodulation performance even at the same power level, and treating interference as AWGN tends to produce a pessimistic estimate of its impact on demodulation. 

To explain why AWGN has a greater impact on LoRa demodulation than BPSK and GFSK, we need to analyse the interference term \( Y_i[k] \).
We begin by applying the principle of stationary phase from  \cite{stationary}, which gives the approximation of the following integral

\begin{equation}
I(\lambda) = \int_a^b g(x)e^{j\lambda f(x)}dx\approx g(c)e^{j\lambda f(c)}e^{\pm j\pi/4}\sqrt{\frac{2\pi}{\lambda |f''(c)|}}.
\label{phase1}
\end{equation}
where \(a\), \(b\), and the function \(f(x)\) are real, \(\lambda\) is a large real parameter. The principle asserts that the dominant contribution to the integral comes from neighborhoods around points where the phase function \(f(x)\) is stationary, i.e., where \(f'(x) = 0\). If \(f(x)\) has a single stationary point at \(x = c\) within the interval \((a, b)\), and if \(f''(c) \neq 0\), then the approximation holds. Moreover, the exponential phase shift is \(+\,j\pi/4\) when \(f''(c)>0\) and \(-\,j\pi/4\) when \(f''(c)<0\).


In our case, we regard \(Y_i[k]\) as the Riemann‐sum approximation to the integral
\begin{equation}
Y_i[k]= \sum_{n=0}^{N-1} i[n]\,e^{\,jF(n;k)}
\;\approx \int_{n=0}^{n=N} i(n)\,e^{j\,F(n;k)}\,dn.
\label{Riemann}
\end{equation}
In this view, the integer index \(n\) is treated as a continuous variable ranging over \([0,N]\). Equivalently, we may introduce the change of variable \(x = \frac{n}{N}, x\in [0,1]\), so that \(n = N\,x\). Substituting \(n = N\,x\) into the expression (\ref{phaseF}) and (\ref{Riemann}), we have
\begin{equation}
\begin{aligned}
F(x;k)
= 2\pi\,N\,\frac{\Delta f}{B}\,x
   \;-\; \pi\,N\,x^2 + \pi\,N\,x
   \;-\; 2\pi\,k\,x,
\end{aligned}
\end{equation}
and
\begin{equation}
Y_i[k]\;\approx\;
N\int_{x=0}^{x=1} \Bigl[i(Nx)\Bigr]\,\exp\Bigl[jN\,\phi(x)\Bigr]\,dx,
\label{phase3}
\end{equation}
where \(\phi(x) \;= \frac{F(n;k)}{N}\).

In this continuous‐integral representation, the factor \(N\) preceding the integral sign arises from \(dn = N\,dx\), and the large parameter \(N\) is now explicit in the exponential term. Thus, we have rewritten the expression for \(Y_i[k]\) in the form of (\ref{phase3}), which closely resembles that of (\ref{phase1}).  By setting
\[
g(x)=i(Nx),\quad f(x)=\phi(x),\quad \lambda=N,
\]
we directly apply formula (\ref{phase1}) to equation (\ref{phase3}).
 The stationary point stands where \(\partial \phi / \partial x = 0\), which yields
\begin{equation}
x_0 \;=\; \frac{\Delta f}{B}-\frac{k}{N}+\frac{1}{2},
\end{equation}
\begin{equation}
n_k = Nx_0 = \frac{N\Delta f}{B}-k+\frac{N}{2}.
\end{equation}
Since \(\partial^2 \phi  / \partial x^2 = -2\pi\) is constant and negative, corresponding to the case \(f''(c)<0\) in (\ref{phase1}), then we have

\begin{align}
Y_i[k] \;\approx\;& 
N\,i[n_k] \cdot e^{jF(n_k;k)} \cdot \sqrt{\frac{2\pi}{N\,|F''(n_k;k)|}} \cdot e^{-j\pi/4} \nonumber \\
=\;& \sqrt{N} \cdot e^{j[F(n_k;k)-\frac{\pi}{4}]} \cdot i[n_k].
\end{align}

Taking the magnitude of the above expression, we obtain
\begin{equation}
|Y_i[k]| \;\approx\; \sqrt{N} \cdot |i[n_k]|,
\label{eq28}
\end{equation}
which shows that the magnitude of  \(Y_i[k]\) approximates a scaled baseband complex envelope of the original sequence \(i[n]\). The corresponding index mapping is
\begin{equation}
k 
\;=\; \frac{N\,\Delta f}{B} + \frac{N}{2} - n_k \quad(\text{mod} \,N).
\label{eq29}
\end{equation}

\begin{figure}[htbp]
\centerline{\includegraphics{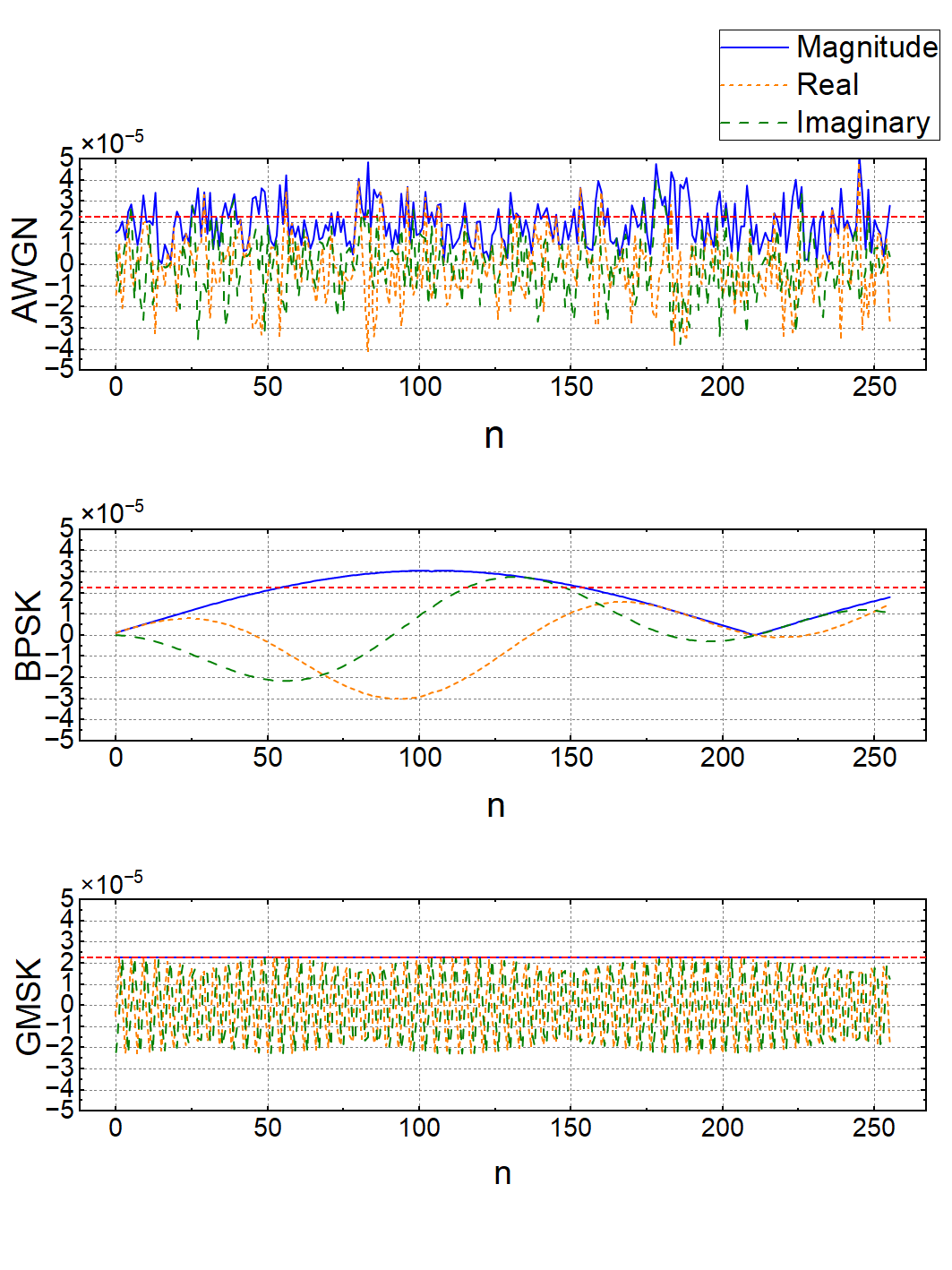}}
\caption{An Example of BPSK, GMSK and AWGN waveforms.}
\label{three_signals}
\end{figure}

Therefore, to analyse the amplitude of \(Y_i[k]\), we can instead study the amplitude of \(i[n]\). When the LoRa and interference signals are only partially time-aligned, some samples of \(i[n]\) become zero, so they contribute nothing to the corresponding bins of \(Y_i[k]\). Fig. \ref{three_signals} shows examples of the complex baseband waveforms \(i[n]\) of the two narrowband signals and the AWGN with the same average power \(P_i\). The red dashed lines indicate the root mean square (RMS) value \(\sqrt{P_i}\). Although these signals share the same average power, their instantaneous amplitudes exhibit different statistical properties, which affect the likelihood of symbol errors.

For AWGN, each sample is a complex Gaussian random variable with zero mean and variance \(P_i/2\) per dimension. As a result, the amplitude follows a Rayleigh distribution with mean \(\sqrt{\pi P_i/4}\), but with a long tail that allows for large values significantly above the RMS value. These peaks can exceed the LoRa signal-induced peak in  \(|Y[k]|\), leading to a higher probability of symbol errors.

For BPSK, the baseband complex envelope becomes zero whenever the encoded bit flips. In order to maintain the same average power \(P_i\), some points must exhibit amplitudes above the RMS value \(\sqrt{P_i}\). This results in moderate instantaneous peaks that can also interfere with LoRa demodulation.

GMSK, in contrast, has a constant-envelope waveform and its amplitude remains fixed at RMS value \(\sqrt{P_i}\). This limited variation reduces the chance of producing peaks that compete with the LoRa signal in the DFT, making GMSK the least disruptive among the three under equal-power conditions.

 \subsection{INR Threshold for Non-zero SER}

In this part, we examine the maximum INR that guarantees a zero symbol error rate for each interference type under various SNR conditions. For a given type of interference and SNR, we initialize the INR at a sufficiently high level so that, over 10,000 Monte Carlo trials, the symbol error rate is greater than zero. We then decrement the INR in fixed steps, repeating the 10,000 trials at each level, until we identify the highest INR at which no symbol errors occur in any trial.

\begin{figure}[htbp]
\centerline{\includegraphics{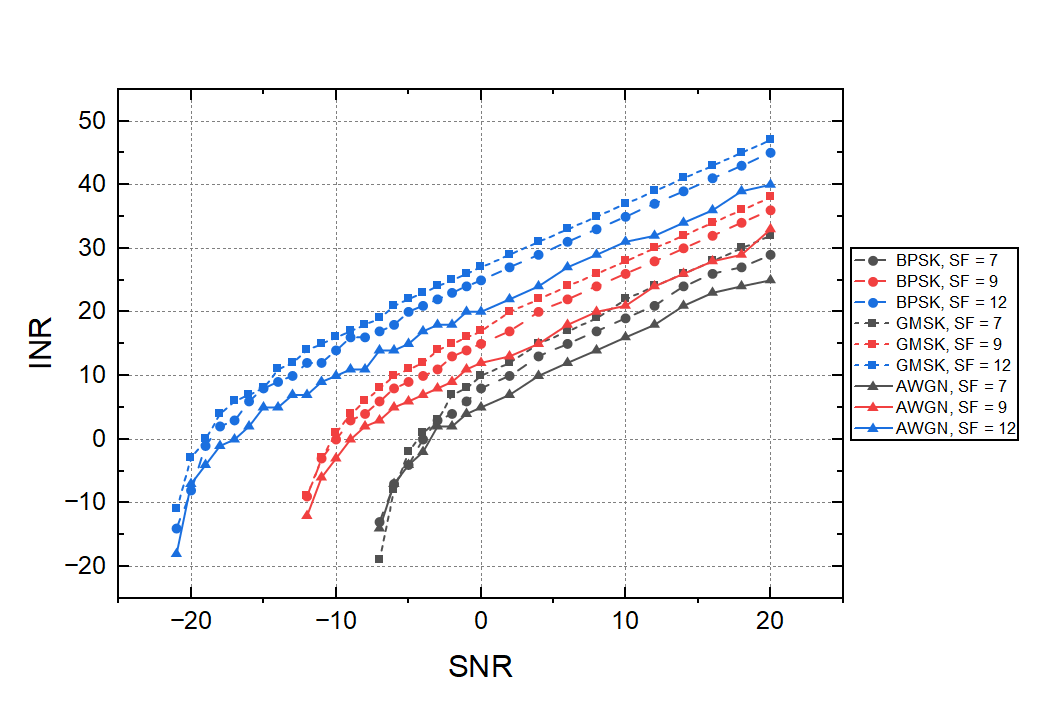}}
\caption{Characteristic Curves of INR-SNR.}
\label{fig5}
\end{figure}

Fig. \ref{fig5} depicts the relation between the maximum INR at the threshold of non-zero SER as a function of SNR for spreading factors 7,9, and 12. The figure can be used to estimate the maximum tolerable INR under given conditions. For instance, when the receiver’s SNR \(=\) 0 dB and GMSK interference is present, an INR not exceeding 10 dB ensures perfect demodulation. 

The INR-SNR characteristic curves exhibit two fundamental behaviours. When the SNR falls to the threshold of \((R_T-N_0-1)\) dB, the maximum acceptable INR approaches negative infinity because the SER inherently exceeds zero due to irreducible noise effects. This results in a vertical asymptote at \text{SNR} \( = (R_T-N_0-1)\) dB. Besides, beyond the transitional region (e.g. SNR \( > (R_T -N_0 + 10)\) dB), a linear dependence between INR and SNR is revealed, which means that at relatively high SNR, further increases in SNR result in a fixed improvement in interference resistance.

To model these behaviours, we propose the parametric function (all quantities in dB)
\[
\mathrm{INR}(\mathrm{SNR})
=
\underbrace{\alpha\,\mathrm{SNR} + \beta}_{\text{High-SNR linear relationship}}
\;+\;
\underbrace{\frac{\gamma}{\mathrm{SNR} - \bigl(R_T - N_{0}-1\bigr)}}_{\text{Low-SNR relationship}}.
\]
where \(\alpha\) defines the linear dependence of INR and SNR at high SNR scenarios, \(\beta\) is the high-SNR intercept, and \(\gamma < 0\) controls the transition steepness. The parameters were fitted using the least squares method with the coefficient of determination \(R^2\) for each fit exceeding 0.99 (see Table~\ref{tab:parameters_sf}).

\setlength{\tabcolsep}{4pt}
\begin{table}[htbp]
\centering
\caption{Fitted parameters for SF7–12 under BPSK, GMSK and AWGN}
\label{tab:parameters_sf}
\begin{tabular}{c|ccc|ccc|ccc}
\hline
\multirow{2}{*}{SF} 
  & \multicolumn{3}{c|}{BPSK} 
  & \multicolumn{3}{c|}{GMSK}
  & \multicolumn{3}{c}{AWGN} \\
\cline{2-10}
  & $\alpha$  & $\beta$ & $\gamma$  
  & $\alpha$ & $\beta$ & $\gamma$
  & $\alpha$ & $\beta$ & $\gamma$ \\
\hline
7  & 1.05 & 9.06 &  -11.11  & 1.08 & 11.36 &  -13.81 & 1.03 & 6.11 &  -7.68    \\
8  & 1.12 & 11.85 &  -10.86  & 1.08& 14.58& -12.95& 0.98& 9.46&  -9.57\\
9  & 1.05& 15.80&  -12.87& 1.04& 18.19& -16.05& 0.99& 12.35&  -13.10\\
10 & 1.03& 19.02& -16.41& 1.05& 20.89& -13.68& 1.00& 15.18&  -11.77\\
11 & 1.01& 22.61& -19.75& 1.02& 24.49& -18.97& 0.96& 18.46&  -18.49\\
12 & 1.02& 25.30&  -13.56& 1.04& 27.04&  -10.84& 1.01& 20.60&  -8.48\\
\hline
\end{tabular}
\end{table}

The fitted parameter \(\alpha\) indicates that in the high-SNR region, an 1 dB increase in SNR yields approximately 1 dB more INR tolerance. This indicates that the power of signal and interference become dominant factors compared to noise floor at high SNR conditions. For a fixed spreading factor, the high-SNR intercept \(\beta\) is the largest for GMSK, smaller for BPSK, and the smallest for AWGN, reflecting their relative interference resilience. Taking SF7 as an example, \(\beta\) equals 11.36 dB under GMSK interference and 9.06 dB under BPSK, showing roughly a 2 dB advantage in LoRa’s resilience to GMSK compared with BPSK at high SNR; from another perspective, for SF7, modeling same-power BPSK and GMSK interference as AWGN (\(\beta =\) 6.11 dB) would underestimate the tolerance by 9.06 \(-\) 6.11 \(=\) 2.95 dB and 11.36 \(-\) 6.11 \(=\) 5.25 dB, respectively.

\section{Conclusion}



In this paper, we employed Monte Carlo simulation to evaluate the SER of LoRa demodulation in the presence of AWGN and interference. We also provided a mathematical analysis of how different narrowband interferers affect the demodulation process. Specifically, we showed that for narrowband interference, its impact on the decision statistic of LoRa demodulation is approximately proportional to the complex envelope of the interferer. Since LoRa selects the bin with the largest magnitude in its decision statistic to demodulate symbols, interferers with larger instantaneous amplitudes yield higher SER under equal-power conditions. This observation explains why AWGN, with its heavy-tailed amplitude distribution, degrades performance more severely than BPSK, which in turn is worse than GMSK. We further showed that the maximum tolerable INR for zero SER at various SNR levels exhibits a two-segment behaviour. In the high-SNR region, each 1 dB increase in SNR yields roughly a 1 dB increase in INR tolerance; in the low-SNR region, however, this tolerance declines rapidly.

Our results indicate that modeling narrowband interference as AWGN leads to overly conservative performance estimates, which highlights the necessity of modulation-specific coexistence evaluations for LoRa networks. We also propose improving carrier sensing strategies for LoRa by first identifying the interferer’s modulation type and then applying adaptive threshold for different modulations to determine channel availability. This approach avoids overly conservative channel estimates, enabling more efficient use of the channel and improving network throughput.

\vfill\eject

\bibliographystyle{unsrt} 
\bibliography{ref}        
\vspace{12pt}
\end{document}